\documentclass[preprint,12pt]{elsarticle}
\usepackage[margin=2.5cm]{geometry}
\usepackage{amssymb}
\journal{arXiv}

\usepackage{graphicx}
\usepackage{amsmath}
\usepackage{amssymb}
\usepackage{tabularx}
\usepackage{newtxmath}
\usepackage{mathbbol}
\usepackage{lineno}
\usepackage[hidelinks]{hyperref}
\usepackage{soul}

\usepackage[dvipsnames]{xcolor}
\usepackage{subfiles}
\usepackage{tikz}
\usepackage{pgfplots}
\usepgfplotslibrary{colorbrewer}
\pgfplotsset{compat=1.18}
\usetikzlibrary{positioning, calc, backgrounds, shapes.geometric, shapes.arrows, arrows.meta, shapes.symbols, decorations.text, fit}
\usetikzlibrary{external}
\usepgfplotslibrary{colormaps}
\usepgfplotslibrary{statistics}
\usepackage[most]{tcolorbox}
\usepackage{hyperref}

\begin{document}

\begin{frontmatter}

\title{Generative Hyperelasticity with Physics-Informed Probabilistic Diffusion Fields}
\author[inst1]{Vahidullah Taç}
\author[inst2]{Manuel K. Rausch}
\author[inst1]{Ilias Bilionis}
\author[inst3]{Francisco Sahli Costabal}
            \ead{corresponding authors: fsc@ing.puc.cl}
\author[inst1,inst4]{Adrian Buganza Tepole}
            \ead{abuganza@purdue.edu}
\affiliation[inst1]{organization={Department of Mechanical Engineering, Purdue University}, 
            city={West Lafayette},
            state={IN},
            country={USA}}
\affiliation[inst2]{organization={Department of Aerospace Engineering and Engineering Mechanics, The University of Texas at Austin}, 
            city={Austin},
            state={TX},
            country={USA}}
\affiliation[inst3]{organization={Department of Mechanical and Metallurgical Engineering and Institute for Biological and Medical Engineering, Pontificia Universidad Catolica de Chile}, 
            city={Santiago},
            country={Chile}}

\affiliation[inst4]{organization={Weldon School of Biomedical Engineering, Purdue University}, 
            city={West Lafayette},
            state={IN},
            country={USA}}

\begin{abstract}
Many natural materials exhibit highly complex, nonlinear, anisotropic, and heterogeneous mechanical properties. Recently, it has been demonstrated that data-driven strain energy functions possess the flexibility to capture the behavior of these complex materials with high accuracy while satisfying physics-based constraints. However, most of these approaches disregard the uncertainty in the estimates and the spatial heterogeneity of these materials. In this work, we leverage recent advances in generative models to address these issues. We use as building block neural ordinary equations (NODE) that -- by construction -- create polyconvex strain energy functions, a key property of realistic hyperelastic material models. We combine this approach with probabilistic diffusion models to generate new samples of strain energy functions. This technique allows us to sample a vector of Gaussian white noise and translate it to NODE parameters thereby representing plausible strain energy functions. We extend our approach to spatially correlated diffusion resulting in heterogeneous material properties for arbitrary geometries. We extensively test our method with synthetic and experimental data on biological tissues and run finite element simulations with various degrees of spatial heterogeneity. We believe this approach is a major step forward including uncertainty in predictive, data-driven models of hyperelasticity.

\end{abstract}

\begin{keyword}
Hyperelasticity \sep generative modeling \sep neural ODEs \sep data-driven modeling \sep heterogeneous materials \sep homogenization
\end{keyword}

\end{frontmatter}

\section{Introduction}
Creating accurate mathematical models of nonlinear mechanical behavior is a central challenge in characterizing complex materials such as skin and other soft tissues. Most biological tissues posses heterogeneous, nonlinear and anisotropic properties \cite{lanir2017multi,jor2013computational}. Machine-learning-based approaches for constitutive modeling have emerged as a potential solution to this challenge \cite{recentadvances, dalDataDrivenHyperelasticityPart2023, eghtesadNNEVPPhysicsInformed2023, rosenkranzComparativeStudyDifferent2023, sacksNeuralNetworkApproaches2022}. Successful prior approaches to capturing soft tissues' constitutive behavior include multi-layer perceptrons (MLP) \cite{liu2020, tac2022data, leng2021, kalina2022automated, fuhgLearningHyperelasticAnisotropy2022}, Gaussian processes (GP) \cite{aggarwalStrainEnergyDensity2023}, input convex neural networks (ICNN) \cite{klein2022polyconvex}, constitutive artificial neural networks (CANN) \cite{linka2023new}, and neural ordinary differential equations (NODE) \cite{tac2022node} (see \cite{tacBenchmarkingPhysicsinformedFrameworks2023a} for a comparative study of ICNNs, CANNs and NODEs for modeling hyperelasticity). Alternatively, stress data have been used to automatically discover material models from a pool of candidate constitutive forms \cite{flaschel2023automated, thakolkaran2022nn, wang2021inference, st.pierreDiscoveringMechanicsArtificial2023}. The rising popularity of data-driven methods in constitutive modeling has also prompted the emergence of auxiliary tools such as finite element solvers oriented around machine learning \cite{pytorchFEA, jaxFEM}.

Data-driven models have not only been highly successful in learning hyperelastic behavior from stress data, they can also be constructed to satisfy physics-based constraints such as polyconvexity  \cite{kleinParametrisedPolyconvexHyperelasticity2023, tac2022node, linka2023new}. This results in physically realistic models and aids in the convergence of partial differential equation solvers, such as the finite element method, which boosts their usability in engineering applications. 
However, most current approaches predict a single, homogeneous material response from a set of experiments. In reality, natural materials show inter- and intra-specimen variability in their mechanical response, i.e., are not homogeneous \cite{matouvs2017review,kouznetsova2001approach,li2019predicting}. Quantifying the resulting uncertainty in biological materials is crucial within the clinical settings \cite{lee2020,stowers2021improving}. This task can be seen as generating plausible samples of material responses, given the limited data that is available. In this regard, the new class of machine learning generative models such as generative adversarial networks (GANs) or diffusion models have the potential to aid in both of these challenges. 

Diffusion is a family of score-based generative models that has been used, with great success, in many applications such as image generation \cite{jolicoeur-martineauAdversarialScoreMatching2020}, audio generation \cite{chenWaveGradEstimatingGradients2020}, protein design \cite{Lee2022.07.13.499967}, and detecting manifolds \cite{pidstrigachScoreBasedGenerativeModels2022}. In generative modeling the task is to approximate the distribution of data, $p(\boldsymbol{\phi})$, given some samples, $\{\boldsymbol{\phi}^i\}_{i=1}^N$, and then generate more samples from this distribution. Score-based generative models achieve this by working with a \emph{score function} instead of the density function, $p(\boldsymbol{\phi})$. The score function is related to the gradient of the probability density as $\nabla_{\boldsymbol{\phi}} \log p(\boldsymbol{\phi})$. The use of score functions results in highly flexible models with tractable and controllable generation process without some of the difficulties of trying to approximate the density function, $p(\boldsymbol{\phi})$ directly \cite{songScoreBasedGenerativeModeling2021, croitoruDiffusionModelsVision2023}. 

We have previously introduced NODEs for modeling the hyperelastic behavior of biological materials \cite{tac2022node}. The models we proposed are built on rigorous continuum-mechanical foundations which enable them to learn the material behavior while satisfying physics-based constraints such as the principle of objectivity, material symmetries, and polyconvexity \textit{a priori}. We use NODEs to model the derivatives of the strain energy density function with respect to invariants of deformation ($I_1, I_2, \cdots$), which means the models are well suited to be used in conjunction with numerical methods such as finite element solvers \cite{tacBenchmarkingPhysicsinformedFrameworks2023a}. We have also extended our methodology to modeling anisotropic finite viscoelasticity \cite{tacDatadrivenAnisotropicFinite2023}. 

\begin{figure*}[h!]
    \centering
    \includegraphics{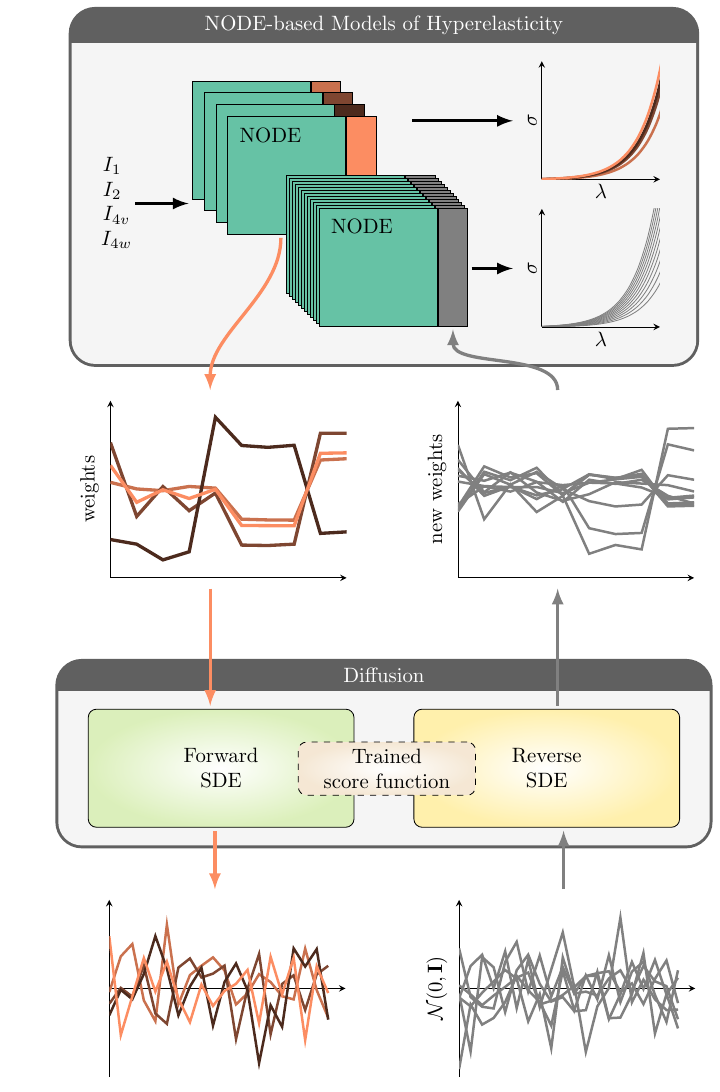}
    \caption{Overview of the method. NODE-based constitutive models yield strain energy density functions $\Psi$ that \textit{a priori} satisfy the conditions of polyconvexity, material symmetries, and objectivity regardless of the parameter values. Given samples of the stress-stretch response $(\boldsymbol{\sigma},\boldsymbol{\lambda})$ for individuals of a population, the goal is to find common weights and biases $\boldsymbol{\varphi}$ shared across samples from that population, as well as individual parameters $\boldsymbol{\phi}^i$. Diffusion is used to estimate the density $p(\boldsymbol{\phi})$ through approximation of the score function $\nabla_{\boldsymbol{\phi}} \log p(\boldsymbol{\phi})$ and thus generate new samples from the population starting from uncorrelated Gaussian noise $\mathcal{N}(\mathbf{0},\mathbf{I})$.   }
    \label{fig_diagram} 
\end{figure*}

In this current study, we construct a robust, principled and fully data-driven generative modeling framework to perform uncertainty quantification to model heterogeneous materials. As backbones, we use NODEs to learn the material response and diffusion to learn the density functions. As shown in Fig. \ref{fig_diagram}, we assume that all but the last layer of every NODE are common across a population of materials (say, skin samples from a group of mice) and the last layer is subject-specific. The subject-specific parameters of the last layer are then used to train a diffusion model based on stochastic differential equations (SDEs) \cite{pidstrigachScoreBasedGenerativeModels2022}. Once trained, the score function can be used in conjunction with the reverse SDE to generate new samples with a Gaussian noise vector as the starting point.

We demonstrate the usability of the framework for synthetic data as well as experimental data from murine skin. The generation process can be conditioned on stress measurements, as well as parameter observations to generate conditional posteriors of the distribution, i.e., $p(\boldsymbol{\phi}|\mathbf{y})$, where $\mathbf{y}$ is a set of observations. Furthermore, we use random fields sampled from zero-mean and unit-variance Gaussian processes to obtain spatially correlated samples, which are useful for generating heterogeneous material responses. We use this to generate a number of heterogeneous samples and perform finite element simulations under various loading conditions. 

\section{Methods}
\subsection{Neural ordinary differential equations}
NODEs are machine learning frameworks such that the output is defined as the solution of an ordinary differential equation (ODE) or a system of ODEs at a given time. 
\begin{align*}
    \frac{d\mathbf{h}(\tau)}{d\tau} = \mathbf{f}_{\text{NODE}}(\mathbf{h}(\tau), \tau, \boldsymbol{\varphi}) \, ,
\end{align*}
where $\tau$ is a pseudo time variable, and $\mathbf{h}$ the variables of interest. The right hand side of the ODE, $\mathbf{f}_{\text{NODE}}(\cdot, \cdot, \boldsymbol{\varphi})$, is a MLP parameterized by $\boldsymbol{\varphi}$. From the fundamentals of ODEs we know that the solution trajectories of ODEs do not intersect, provided that the right hand side is Lipschitz continuous. For the scalar variable case, $h(\tau)$, this implies that for trajectories $h_1$ and $h_2$, the following holds
\begin{align*}
    h_1(0) \geq h_2(0) \Longleftrightarrow h_1(1) \geq h_2(1)\, ,
\end{align*}
which means that the input-output map of a scalar NODE is monotonic. We employ this property of NODEs to construct polyconvex strain energy density functions as we explain in the next subsection.

\subsection{Polyconvex strain energy density functions with NODEs}
The construction of polyconvex data-driven strain energy density functions is based on our previous publication \cite{tac2022node} and briefly reported here for completeness. We propose strain energy density functions of the form 

\begin{multline}
    \Psi(\mathbf{F}) = \Psi_{I_1}(I_1)+   \Psi_{I_2}(I_2)+   \Psi_{I_{4v}}(I_{4v}) +\Psi_{I_{4w}}(I_{4w}) + \sum_{j > i} \Psi_{I_i,I_j} \left(\alpha_{ij} I_i+ (1-\alpha_{ij})I_j\right)+ \Psi_{J}(J)\, ,
    \label{eq_strain energy density function_general}
\end{multline}
where $I_1, I_2, I_{4v}$ and $I_{4w}$ are invariants of the right Cauchy-Green deformation tensor, $\mathbf{C}=\mathbf{F}^\top \mathbf{F}$, and $J=\det \mathbf{F}$. We have shown that in order to preserve polyconvexity of the strain energy density function each of the $\Psi_{I_i}$ and $\Psi_{I_i, I_j}$ needs to be convex non-decreasing. This is equivalent to monotonicity of the derivatives, $d\Psi_{I_i}/d I_i$ with $d\Psi_{I_i}/d I_i \geq 0$ in the domain of $I_i$. We have also shown that NODEs are monotonic functions and can be made non-negative with minor modifications to the neural network architecture \cite{tac2022node, tacDatadrivenAnisotropicFinite2023}. As a result, modeling the derivative functions $d\Psi_{I_i}/d I_i$ with NODEs guarantees polyconvexity of the strain energy density function. 

The architecture of the NODEs can be adjusted depending on nonlinearity and complexity of the material response. We have previously explored the trade-off between number of parameters and accuracy of NODE material models \cite{tacBenchmarkingPhysicsinformedFrameworks2023a}. 

\subsection{Score-based generative modeling \& diffusion}

In generative modeling, the task is to find the model distribution $\hat{p}(\boldsymbol{\phi})$ that best approximates the data distribution $p(\boldsymbol{\phi})$ given some samples $\{\boldsymbol{\phi}_0^{i}\}_{i=1}^N $, where the subscript $0$ is used to imply $t=0$ in the diffusion process. For a given  density function $p(\boldsymbol{\phi})$, a score-based generative model employs two SDEs. 
The first one is called the \emph{forward SDE}, and has the It\^{o} form
\begin{align}
    d \boldsymbol{\phi}_t &= \mathbf{f}(\boldsymbol{\phi}_t,t) dt + \mathbf{g}(t) d \mathbf{B}_t \, , \label{eq:fwdSDE_general}
\end{align}
for $t$ between $0$ and $T$, and initial conditions $\boldsymbol{\phi}_0$ sampled from  $p(\boldsymbol{\phi})$. Here $\mathbf{B}_t$ is m-dimensional Brownian motion (or Wiener process) and $\mathbf{f}(\boldsymbol{\phi}_t,t)$ and $\mathbf{g}(t)$ are functions that calculate the drift and diffusion coefficients, respectively. The drift coefficient is designed such that it gradually turns the data $\boldsymbol{\phi}_0$ into noise, while the diffusion coefficient controls the amount of Gaussian noise added in each step. 

The second SDE, known as the \emph{reverse SDE} \cite{songScoreBasedGenerativeModeling2021}, it runs for $t$ from $T$ to $0$, and it has the form
\begin{equation}
    d \boldsymbol{\phi}_t = \left[\mathbf{g}^2(t) \nabla_{\boldsymbol{\phi}} \log p_t(\boldsymbol{\phi}) - \mathbf{f}(\boldsymbol{\phi}_t,t) \right] dt + \mathbf{g}(t) d \hat{\mathbf{B}}_t,
\end{equation}
where $\hat{\mathbf{B}}_t$ represents the Brownian motion when time is reversed and the quantity $\nabla_{\boldsymbol{\phi}} \log p_t(\boldsymbol{\phi})$ is known as the \emph{score function}. Note that this introduces the density $p_t(\boldsymbol{\phi}_t)$, which is related to the forward SDE as will be explained shortly. It can be shown that if we start with Gaussian noise ($\boldsymbol{\phi}_T \sim \mathcal{N}(\mathbf{0},\mathbf{I})$ with $\mathbf{I}$ the identify matrix and $\mathcal{N}(\boldsymbol{\mu},\boldsymbol{\Sigma})$ the normal distribution with mean $\boldsymbol{\mu}$ and covariance $\boldsymbol{\Sigma}$), the reverse SDE recovers the original data by removing the drift responsible for the destruction of the data. 

There are a number of popular choices for the forward and the associated reverse SDE used in the literature, such as Brownian motion, critically damped Langevin dynamics, and the Ornstein-Uhlenbeck (OU) process \cite{pidstrigachScoreBasedGenerativeModels2022}. In this study, we use the scaled OU process for the forward and reverse SDEs, given as
\begin{align}
    d \boldsymbol{\phi}_t &= -\frac{1}{2} \beta(t) \boldsymbol{\phi}_t dt + \sqrt{\beta(t)}d \mathbf{B}_t, &\text{(forward SDE)}
    \\
    d \boldsymbol{\phi}_t &= \frac{1}{2} \beta(t) \boldsymbol{\phi}_t dt + \beta(t) \nabla \log p_{t} (\boldsymbol{\phi}_t) dt +\sqrt{t \beta(t)} d\hat{\mathbf{B}}_t. &\text{(reverse SDE)}
    \label{eq:reverseOU}
\end{align}
One of the advantages of the OU process is its simple form. Since both dispersion and drift functions are linear in $\boldsymbol{\phi}$, the solution of the forward SDE is given in closed form as a normally distributed random variable. The marginal of $\boldsymbol{\phi}_t$ conditioned on the observation of the starting point $\boldsymbol{\phi}_0$ is a normal distribution 
\begin{equation}
    p_{t|0} (\boldsymbol{\phi}_t|\boldsymbol{\phi}_0) = \mathcal{N} \left( \boldsymbol{\phi}_t \, | \, \mu(t)\boldsymbol{\phi}_0, \Sigma(t) \mathbf{I} \right) \, ,
    \label{eq:pt0}
\end{equation}
where the mean $\mu(t)$ and variance $\Sigma(t)$ depend on the scaling $\beta(t)$. For example, for the standard OU process with no scaling, i.e. $\beta=1$,
\begin{equation}
    p_{t|0} (\boldsymbol{\phi}_t|\boldsymbol{\phi}_0) = \mathcal{N} \left(\boldsymbol{\phi}_t|  \exp(-t/2) \boldsymbol{\phi}_0, (1-\exp(-t)) \mathbf{I} \right) \, .
    \label{eq:pt0_OU}
\end{equation}
Here we follow the recommendations from \cite{karras2022elucidating,pidstrigach2022scorebased} and choose the scaling 
\begin{equation*}
    \beta(t) = \beta_{\min} + t(\beta_{\max}-\beta_{\min})\, ,
\end{equation*}
with hyper-parameters $\beta_{\min}=0.001$, $\beta_{\max}=3$ \cite{karras2022elucidating}, which leads to mean and variance of the conditional distribution (\ref{eq:pt0}),
\begin{align*}
    \mu(t) &= \exp(-\alpha(t)/2), \\
    \Sigma(t) &= 1-\exp(-\alpha(t)) \, ,
\end{align*}
with 
\begin{equation*}
    \alpha(t) = \int_0^t \beta(s)ds=\beta_{\min} t + \frac{1}{2} t^2 (\beta_{\max}-\beta_{\min})\, .
\end{equation*}

The importance of the marginal Eq. (\ref{eq:pt0_OU}) will become evident later. As an initial motivation, note that the density $p_t(\boldsymbol{\phi}_t)$ and therefore the score function $\nabla_{\boldsymbol{\phi}} \log p_t(\boldsymbol{\phi})$ is unknown. In score-based generative models, the score function is approximated by a neural network $s_{\boldsymbol{\theta}}(\boldsymbol{\phi},t) \approx \nabla_{\boldsymbol{\phi}} \log p_t(\boldsymbol{\phi})$ parameterized by weights and biases $\boldsymbol{\theta}$ in a process known as \emph{score matching}. 

\subsubsection{Score matching}
We would like to minimize the training loss defined as
\begin{equation*}
    L(\boldsymbol{\theta}) = \int_0^T \mathbb{E}_{p_t} \left[||\nabla_{\boldsymbol{\phi}} \log p_t(\boldsymbol{\phi}) - s_{\boldsymbol{\theta}}(\boldsymbol{\phi},t)||^2 \right] dt,
\end{equation*}
where $\mathbb{E}_{p_t}$ denotes the expectation over $p_t(\boldsymbol{\phi})$. However, the density function $p_t(\boldsymbol{\phi}_t)$ is unknown. Instead we can approximate $p_t(\boldsymbol{\phi}_t)$ from the data. First, note that $p_t(\boldsymbol{\phi}_t)$ can be obtained by marginalizing the conditional on the initial distribution of the forward SDE

\begin{equation*}
    p_t(\boldsymbol{\phi}_t) = \int p_{t|0}(\boldsymbol{\phi}|\boldsymbol{\phi}_0) p(\boldsymbol{\phi}_0) d\boldsymbol{\phi}_0 \, .
\end{equation*}
We do not have the distribution of the data, however, we can approximate it based on the available samples,

\begin{equation*}
    p_0(\boldsymbol{\phi}_0)  \approx \hat{p}_0(\boldsymbol{\phi}_0) = \sum_{i=1}^N \delta(\boldsymbol{\phi}-\boldsymbol{\phi}_0^i)\, .
\end{equation*}

With this approximation of $p_0(\boldsymbol{\phi}_0)$, we get our approximation of $p_t(\boldsymbol{\phi}_t)$ which we denote $ \hat{p}_t(\boldsymbol{\phi}_t)$,
\begin{equation}
    \hat{p}_t(\boldsymbol{\phi}_t) = \mathbb{E}_{\hat{p}_{0} }[p_{t|0}(\boldsymbol{\phi}|\boldsymbol{\phi}_0)] = \frac{1}{N} \sum_{i=1}^N p_{t|0} \left(\boldsymbol{\phi}|\boldsymbol{\phi}^{i}_0\right),
    \label{eq:phat}
\end{equation}
where $p_{t|0} (\boldsymbol{\phi}|\boldsymbol{\phi}^{i}_0)$ are the Gaussians defined in (\ref{eq:pt0_OU}) evaluated at the points $\boldsymbol{\phi}_0^i$ \cite{vincent2011connection}, for which the logarithm and gradient can be easily computed. 
Substituting this, we obtain the surrogate loss as
\begin{equation}
    \hat{L}(\boldsymbol{\theta}) = \int_0^T \mathbb{E}_{\hat{p}_t}\left[||\nabla_{\boldsymbol{\phi}} \log \hat{p}_t(\boldsymbol{\phi}) - s_{\boldsymbol{\theta}}(\boldsymbol{\phi},t)||^2\right] dt.
    \label{eq:Loss}
\end{equation}

The score function $s_{\boldsymbol{\theta}}$ is captured with an MLP parameterized by $\boldsymbol{\theta}$. To evaluate Eq. (\ref{eq:Loss}), samples  from the uniform distribution $t\sim \mathcal{U}(0,1)$ are drawn, as well as samples from Eq. (\ref{eq:phat}) to obtain the expectations. We use 4 layers with 256 neurons each and train the score for different number of epochs depending on the dimensionality of the data, in the range [500,2000]. The training is conducted on an Apple M1 Pro CPU using JAX high performance numerical library in Python and the Adam optimizer with a step size of $1\times 10^{-5}$ and parameters $\beta_1=0.9$ and $\beta_2=0.999$ (as named by JAX). Training batch sizes vary from 10 to 200 depending on the sample size. Training time is on the order of minutes.


\subsubsection{Sample generation}

Once trained, the score $s_{\boldsymbol{\theta}}$ is used in place of $\nabla_{\boldsymbol{\phi}} \log p_{t} (\boldsymbol{\phi}_t)$ in the reverse SDE Eq. (\ref{eq:reverseOU}) with initial conditions $\boldsymbol{\phi}_T \sim \mathcal{N}(\mathbf{0},\mathbf{I})$. For the integration of the reverse SDE we perform 1000 time steps of the Euler–Maruyama scheme with a time step of $\Delta t = 1\times 10^{-3}$ from $t=0$ to $T=1$.

\subsubsection{Conditional diffusion}
We can condition the generation process to obtain samples that are closer to a given measurement. The score function can be obtained from the density by differentiating its logarithm and the density can be recovered up to a constant by integrating the score function. This means working with the score function is equivalent to working with the density function. This allows us to use Bayes' rule to condition the generation process on a set of observations $\mathbf{y}$:
\begin{align*}
    p(\boldsymbol{\phi}|\mathbf{y}) = \frac{p(\mathbf{y}|\boldsymbol{\phi})p(\boldsymbol{\phi})}{p(\mathbf{y})}\, ,
\end{align*}
or -- in terms of the score function -- \cite{songScoreBasedGenerativeModeling2021},
\begin{equation*}
    \nabla_{\boldsymbol{\phi}} \log p(\boldsymbol{\phi}|\mathbf{y}) = \underbrace{\nabla_{\boldsymbol{\phi}} \log p(\mathbf{y}|\boldsymbol{\phi})}_{\text{Score of the likelihood}} + \underbrace{\nabla_{\boldsymbol{\phi}}\log p(\boldsymbol{\phi})}_{\text{The trained score, } \approx s_{\boldsymbol{\theta}}(\boldsymbol{\phi},t)}.
\end{equation*}
This new score function can be plugged into the reverse SDE after training to obtain samples conditioned on observations $\mathbf{y}$.

For example, consider $\boldsymbol{\phi}$ to be the parameters of a NODE. Let $\mathbf{y}\in \mathbb{R}^{n_{\text{obs}}}$ be the observation of some quantities of interest, e.g. the stress, and $\hat{\boldsymbol{\sigma}} (\boldsymbol{\phi}) : \mathbb{R}^N \to \mathbb{R}^{n_\text{obs}}$ the material model connecting the variables $\boldsymbol{\phi}$ to the quantity of interest. Assuming a Gaussian likelihood with noise $\varsigma$, the corresponding score is

\begin{equation}
    \nabla_{\boldsymbol{\phi}} \log p(\mathbf{y} | \boldsymbol{\phi}) =  -\frac{1}{2 \varsigma^2} \sum_{i=1}^{n_{\text{obs}}} \left((y_i -\hat{\sigma}_i(\boldsymbol{\phi})) \nabla_{\boldsymbol{\phi}} \hat{\sigma}_i \right).    
    \label{eq_conditional_diffusion_s}
\end{equation}

 The reverse SDE is analogous to (\ref{eq:reverseOU}) but with the additional score
 
\begin{equation}
    d \boldsymbol{\phi}_t = \frac{1}{2} \beta(t) \boldsymbol{\phi}_t dt +\beta(t) \left(\nabla_{\boldsymbol{\phi}} \log p_t(\boldsymbol{y}|\boldsymbol{\phi_t}) + \nabla_{\boldsymbol{\phi}} \log p_{t} (\boldsymbol{\phi}_t) \right) dt + \sqrt{t \beta(t)} d\hat{\mathbf{B}}_t \, .
    \label{eq:conditional_SDE}
\end{equation}

Note that while it is possible to use the likelihood as in (\ref{eq_conditional_diffusion_s}), plugging in the current value of the variable $\boldsymbol{\phi}_t$, a time dependent likelihood can be designed as proposed in \cite{chung2022improving,chung2022diffusion}. 

\subsection{Generative hyperelasticity}

As shown in Fig.~\ref{fig_diagram}, the goal of the proposed framework is to generate constitutive models for a population, e.g., the mechanical behavior of murine skin. The material models are defined by strain energy density functions, which are functions of deformation. We want to sample these functions from a distribution that characterizes the materials from a population such that they satisfy the desired polyconvexity, objectivity, and material symmetries. Our approach is akin to hypernetwork or latent space approaches in \cite{du2021learning,dupont2022data}, with the difference that the strain energy density function is represented in terms of NODEs in Eq. (\ref{eq_strain energy density function_general})  guarantees that any sample of NODE parameters already satisfies the desired constraints for the strain energy \textit{a priori}. We split the weights and biases of the NODEs into a subset $\boldsymbol{\varphi}$ which is shared among all individuals of the population, and a subset $\boldsymbol{\phi}^i$, the weights of the last layer of the NODEs, which is specific to an individual $i$ (see Fig. \ref{fig_diagram}). Our explicit notation for the strain energy density function combining shared and individual parameters is $\Psi_{\boldsymbol{\varphi},\boldsymbol{\phi}^i}(\mathbf{F})$. Our goal is then to find the density over the parameters $p(\boldsymbol{\phi})$ that characterizes the population. 
 
Consider stress-stretch data from individuals, $\boldsymbol{\sigma}^{i}(\boldsymbol{\lambda})$ where $i=1,...,M$ are the individuals of the population for which we have observations of the stress for different deformations $\mathbf{\lambda}$. The first goal is to find the population $\boldsymbol{\varphi}$ and individual $\boldsymbol{\phi}^{i}$ that can capture the data $\boldsymbol{\sigma}^{i}(\boldsymbol{\lambda})$. We achieve this goal by minimizing a standard loss that computes the mean average error between the data and the predicted stresses from the model $\boldsymbol{\sigma}_{\text{pred}}^{i}(\boldsymbol{\lambda}|\boldsymbol{\varphi},\boldsymbol{\phi}^{i})$. Then, given the samples $\boldsymbol{\phi}^{i}$, we use the probabilistic diffusion methods outlined above to estimate the density $\hat{p}(\boldsymbol{\phi})$ which allows us to generate new models $\Psi_{\boldsymbol{\varphi},\boldsymbol{\phi}}$ from the population and derived quantities of interest such as stress density $\hat{p}(\boldsymbol{\sigma})$.

\subsection{Heterogeneous material fields}

Many materials are heterogeneous, in other words, the material response depends on the spatial coordinates, such that the strain energy density function can be written as  $\Psi(\mathbf{{F}},\mathbf{x})$. We leverage the generative hyperelastic models from the last section in order to generate not just point estimates of the material response according to the fitted distribution, but spatially heterogeneous material fields. The key advantage of diffusion models is that we start from the uncorrelated Gaussian noise $\boldsymbol{\phi}_T \sim \mathcal{N}(\mathbf{0},\mathbf{I})$. Thus, to produce spatially correlated fields $\boldsymbol{\phi}(\mathbf{x})$ we instead start by sampling a GP $\boldsymbol{\phi}_T(\mathbf{x}) \sim \left(0,k(\mathbf{x},\mathbf{x}';\vartheta)\right)$ where $k$ is a covariance function such that it has unit variance and correlation lengths $\vartheta$, 

\begin{equation}
    k(\mathbf{x},\mathbf{x}';\vartheta) =  \exp\left\{-\sum_{i=1}^d\frac{\left(x_i - x_i'\right)^2}{2\vartheta_i^2}\right\}, 
    \label{eq:kernel}
\end{equation}
with $\vartheta_i$ denotes the length scales of the correlations for each coordinate of the points  $\mathbf{x}\in \mathbb{R}^d$. Thus, given a trained score function, spatially correlated fields $\Psi_{\boldsymbol{\varphi},\boldsymbol{\phi}}(\mathbf{x})$ of material properties can be generated by the Euler-Maruyama scheme


\begin{equation}
    \boldsymbol{\phi}_{t-\Delta t}(\mathbf{x}) =\boldsymbol{\phi}_{t}(\mathbf{x}) + \frac{1}{2} \boldsymbol{\phi}_t(\mathbf{x}) \beta(t) \Delta t + \nabla \log p_{t} (\boldsymbol{\phi}_t(\mathbf{x})) \beta(t) \Delta t + \hat{\mathbf{Z}}_t(\mathbf{x})  \sqrt{\Delta t \beta(t)}
    \label{eq_Euler_stochastic_PDE}
\end{equation}
with time step $\Delta t=0.001$ and $\hat{\mathbf{Z}}_t(\mathbf{x})$ a GP with zero mean, kernel (\ref{eq:kernel}), and the same length scales $\vartheta$. For complex geometries, sampling GPs relies on an expansion using the eigenfunction of the Laplace operator as in \cite{elhag2023manifold,borovitskiy2020matern,gander2022fast}. In this case, the eigenfunctions are approximated with finite elements. With this approximation we recover a Mat\'ern kernel 5/2, for which we also set the variance to one and change the length scale. Briefly, given a manifold $\mathcal{B}$, 
the  eigenvalues of the Laplace-Beltrami operator are obtained by solving 

\begin{equation*}
    \Delta e_i(\mathbf{x}) = \lambda_i e_i(\mathbf{x})
\end{equation*}
where $\lambda_i$ are the eigenvalues corresponding to the eigenfunctions $e_i(\mathbf{x})$. For a domain discretized with finite elements, the eigenvalue problem can be computed with standard finite element techniques \cite{hughes2012finite}. Given $\lambda_i,e_i(\mathbf{x})$, the kernel for the manifold can be expressed as 

\begin{equation}
    k_\mathcal{B}(\mathbf{x},\mathbf{x}';C_m, \nu, \kappa) = \frac{1}{C_m} \sum_{n=0}^\infty \left(\frac{2\nu}{\vartheta}^2+\lambda_n\right)^{-\nu - d/2} e_n(\mathbf{x}) e_n(\mathbf{x}') \, ,
    \label{eq:matern}
\end{equation}
where $\nu, \vartheta$ are the hyperparameters of the kernel, $d$ is the dimension of the manifold, and $C_m$ is a normalizing constant to ensure that samples of the Gaussian process on $\mathcal{B}$ have unit variance. The constant can be determined by $\left(V_{\mathcal{B}}^{-1} \int_\mathcal{B} k_\mathcal{B}(\mathbf{x},\mathbf{x})\right)^{1/2}=1$ with $V_\mathcal{B}$ the volume (or area) of the manifold. The length scale is controlled by the hyperparameter $\vartheta$ \cite{borovitskiy2020matern}. 

\subsection{Model calibration and verification}
Strain energy density functions of the form \eqref{eq_strain energy density function_general} encompass a large ansatz by including potential interactions between all the different invariants. However, in practice we have found that a smaller ansatz can describe most materials extremely well. In this study we use assume that the materials are incompressible ($J=1, d\Psi_J/dJ=0$) and use the following formulation for isotropic and anisotropic materials
\begin{alignat*}{3}
    \Psi(\mathbf{F}) = &\Psi_{I_1}(I_1) + \Psi_{I_2}(I_2)  &\text{(Isotropic, 2 NODEs)} \phantom{\, .} 
    \\
    \Psi(\mathbf{F}) = &\Psi_{I_1}(I_1) + \Psi_{I_2}(I_2) &
    \\
    &+ \Psi_{I_1,I_{4v}}(\alpha_1 I_1 + (1-\alpha_1)I_{4v}) &
    \\
    &+ \Psi_{I_1,I_{4v}}(\alpha_2 I_1+ (1-\alpha_2)I_{4w}) &
    \\
    &+ \Psi_{I_{4v},I_{4w}}(\alpha_3 I_{4v}+ (1-\alpha_3)I_{4w}) \quad \quad &\text{(Anisotropic, 5 NODEs)} \, .
\end{alignat*}
All but the last layer of the neural networks in each of the NODEs are common across the population and are not used in the generative modeling process, whereas the last layer is individual-specific. 

For synthetic data generation, we use strain energy functions of the form
\begin{align}
    \label{eq_Psi_MN}
    \Psi_{MN} = \frac{k_1}{k_2} (\exp(k_2(I_1-3))-1) + \mu(I_1-3)
\end{align}
often referred to as the May-Newman model \cite{may1998constitutive}.

For experimental data, we use biaxial stress ($\sigma_{xx},\sigma_{yy}$) - stretch ($\lambda_x,\lambda_y$) data from murine skin from \cite{meador2020}. The data are collected under off-X ($\lambda_y=\lambda, \lambda_x=\sqrt{\lambda}$), off-Y ($\lambda_x=\lambda, \lambda_y=\sqrt{\lambda}$) and equibiaxial ($\lambda_y=\lambda_x=\lambda$) loading conditions from the ventral and dorsal regions of 15 different mice.

\section{Results}
\subsection{Synthetic data evaluation}
\begin{figure*}[h!]
    \centering
    
    \includegraphics{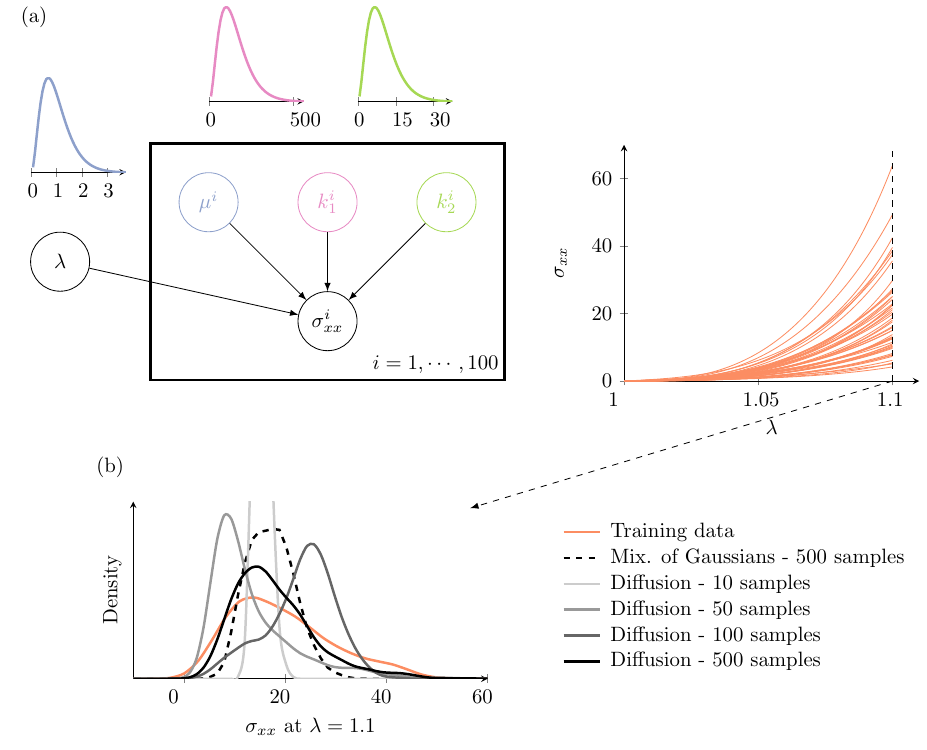}
    \caption{Synthetic data example. An analytical material model with three parameters is used and samples from the joint distribution of the parameters of the material model $\mu,k_1,k_2$ are evaluated at different deformations $\lambda_x = \lambda_y = \lambda$ to get the corresponding stress $\boldsymbol{\sigma}$ for a total of 10,000 individuals from this population, the first 500 of which is used for training the model}. The NODE model is trained on the $(\boldsymbol{\sigma},\lambda)$  data to obtain common parameters $\boldsymbol{\varphi}$ as well individual parameters $\boldsymbol{\phi}^i$. Score matching diffusion is used to estimate $\hat{p}(\boldsymbol{\phi})$ which enables estimation of distribution of quantities of interest such as the stress at a particular deformation $\hat{p}(\sigma_{xx}|\lambda=1.1)$. The predicted distribution of stress matches closely the true distribution as the number of samples from the population are increased from 5 to 100, outperforming traditional density estimation methods.
    \label{fig_analytical} 
\end{figure*}

We first test the generative modeling framework with synthetic data generated from a strain energy density function of the form \eqref{eq_Psi_MN}. We specifically choose an exponential strain energy density function to challenge the framework. The three parameters of strain energy density function are sampled from asymmetric gamma distributions and 10000 stress-stretch curves were generated from this population to approximate the distribution of stress. The distribution of the underlying material parameters and the first 100 of the resulting stress responses are shown in Fig. \ref{fig_analytical} (a). We use the distribution of stresses at $\lambda=1.1$ in equibiaxial loading, $p(\sigma_{xx}|\lambda=1.1)$, to benchmark the performance of the diffusion model when trained with 5, 10, 50, 100, and 500 samples from the population (Fig. \ref{fig_analytical} (b)). Namely, NODE based models were trained to the stress-stretch data to obtain the common parameters $\boldsymbol{\varphi}$ as well as the individual sample parameters $\boldsymbol{\phi}^i$ for samples $i$ of the population. The diffusion scheme was then used to estimate the density $\hat{p}(\boldsymbol{\phi})$ out of the individual NODE models, and to then produce the estimate of $\hat{p}(\sigma_{xx}|\lambda=1.1)$ by generating new samples from  $\hat{p}(\boldsymbol{\phi})$ and evaluating the NODE-based strain energy density function (\ref{eq_strain energy density function_general}). The results of diffusion match the training data increasingly well as the number of samples from the population is increased from 5 to 500. As a benchmark we perform the same study with a classic probability estimation method, the mixture of Gaussians. This approach using all 500 observations performs fairly well in this simple, isotropic case but not in the anisotropic case when the number of parameters is increased (See Supplement Fig. 2). Even though the mixture of Gaussians can reasonably estimate $p(\sigma_{xx}|\lambda=1.1)$, it is unable to capture the skewness of the distribution and fails when compared to the probabilistic diffusion model.

\begin{figure*}[h!]
    \centering
    \includegraphics[width=11cm]{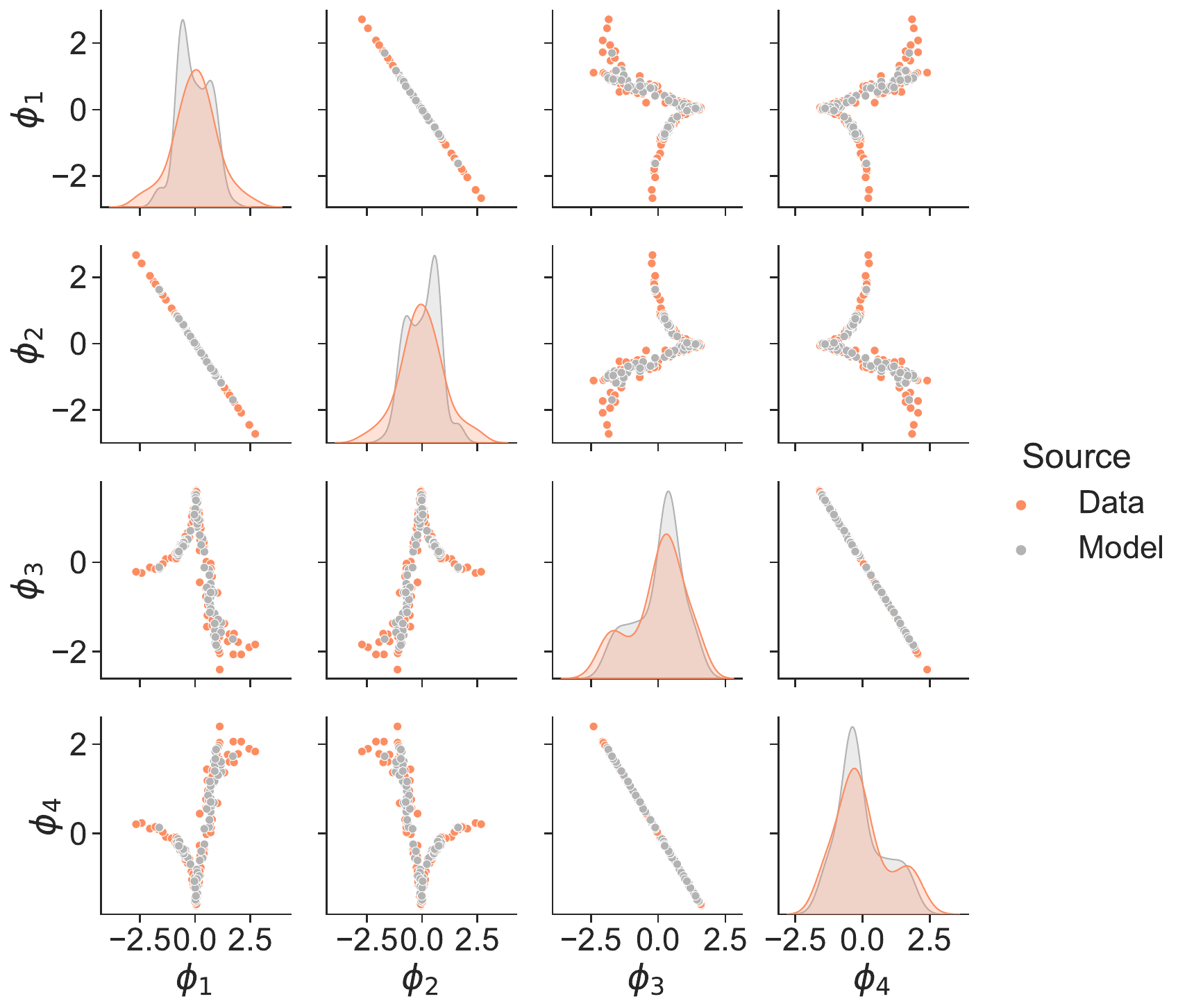}
    \caption{Pair plot to visualize the distribution of the NODE parameters $\boldsymbol{\phi}$. 100 samples of parameters $\boldsymbol{\phi}^i$ were used to train the diffusion generative model. Comparison between a direct kernel density estimation of $p(\boldsymbol{\phi})$ and the density estimation from diffusion shows that the score matching model can capture the distribution accurately.}
    \label{fig_pairplot} 
\end{figure*}

Fig. \ref{fig_pairplot} shows pair plots of the distribution of the NODE parameters. This figure is to verify  that the density estimation with diffusion is equivalent to the kernel density estimation directly applied to 100 $\boldsymbol{\phi}^i$ samples of NODE parameters. As can be seen, the diffusion generative model has no problem in capturing the joint probability of the parameters $\boldsymbol{\phi}$, however it has a tendency to concentrate more avidly towards the center of the distribution. This figure also shows that the joint probability of the NODE parameters is not necessarily Gaussian, which explains why the mixture of Gaussians from Fig. \ref{fig_analytical} has difficulty estimating the material behavior of the analytical case, and fails with the anisotropic skin data (see Supplement).

As the size of the neural networks grow, the NODEs become increasingly more flexible, allowing them to capture the training data more closely. However, this could come at the cost of increased dimensionality for the density estimation problem. We checked whether the increased dimensionality of $\boldsymbol{\phi}$ from 4 to 18 had an effect on the density estimation for the analytical problem. We found that there was no decrease in performance estimating $\hat{p}(\boldsymbol{\phi})$ for this example (see Supplemental Fig. 1). 


\subsection{Experimental data characterization}

\begin{figure*}[h!]
    \centering
    \includegraphics[width=.9\textwidth]{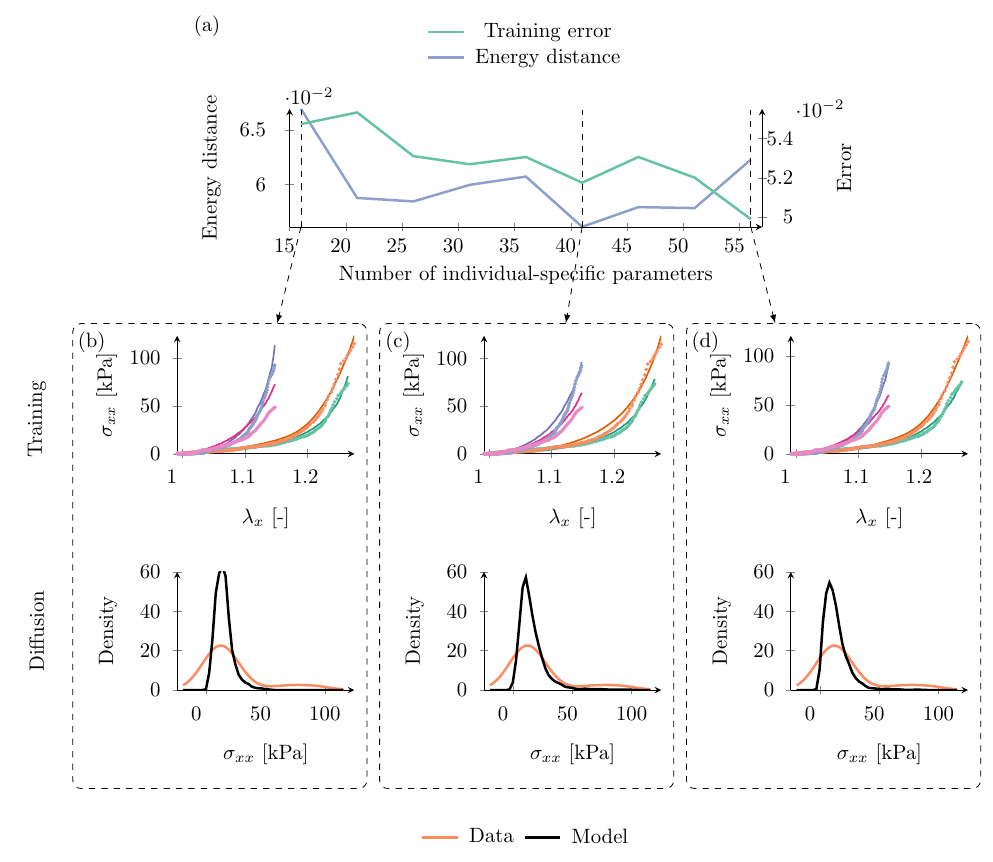}
    \caption{Estimation of the distribution of murine skin mechanical behavior using NODEs and diffusion probabilistic models. Stress-strain data $(\boldsymbol{\sigma},\boldsymbol{\lambda})$ from 15 mice was used to train NODE-based strain energy density functions by minimizing the training loss (a). The parameter samples underlying the material response, $\boldsymbol{\phi}^i$, were then used to estimate the density $\hat{p}(\boldsymbol{\phi})$ using diffusion. The total variation distance between the empirical distribution of a quantity of interest (stress at a given stretch) estimated directly from the experimental data and the corresponding distribution generated by sampling $\hat{p}(\boldsymbol{\phi})$ and evaluating the NODE-based strain energy density function was used as a metric of model quality (a). There is a trade-off between increasing model complexity, training loss, and total variation distance (b,c,d). Solid lines correspond to the NODE predictions and dotted lines correspond the data.}
    \label{fig_mice_w_sensitivity} 
\end{figure*}

Next we train the model with experimental data from murine skin. Based on our results with the synthetic data, for the murine skin dataset we start by determining the right number of subject-specific parameters (Fig. \ref{fig_mice_w_sensitivity}). As the number of parameters is increased the training loss decreases whereas the graph of energy distance displays a valley with a minimum at 31 parameters. The energy distance \cite{rizzoEnergyDistance2016}, is a measure of the discrepancy between two distributions that satisfies the usual requirements of a distance function. See the Supplement for additional information on the definition of the energy distance. In this case, the energy distance is computed between the empirical distribution of the quantity of interest based directly on the data (stress at a particular deformation) and the corresponding distribution of the stress generated by sampling $\hat{p}(\boldsymbol{\phi})$ and evaluating the NODE model at the desired deformation. Even though the energy distance is not computed with respect to the actual distribution, because it is not available, this is still a meaningful test that gives confidence on a good estimation of the underlying parameterization of the material model. Based on Fig. \ref{fig_mice_w_sensitivity}, we determined that 31 parameters was a good choice for the subsequent analyses.

\begin{figure*}[h!]
    \centering
    \includegraphics[width=0.9\textwidth]{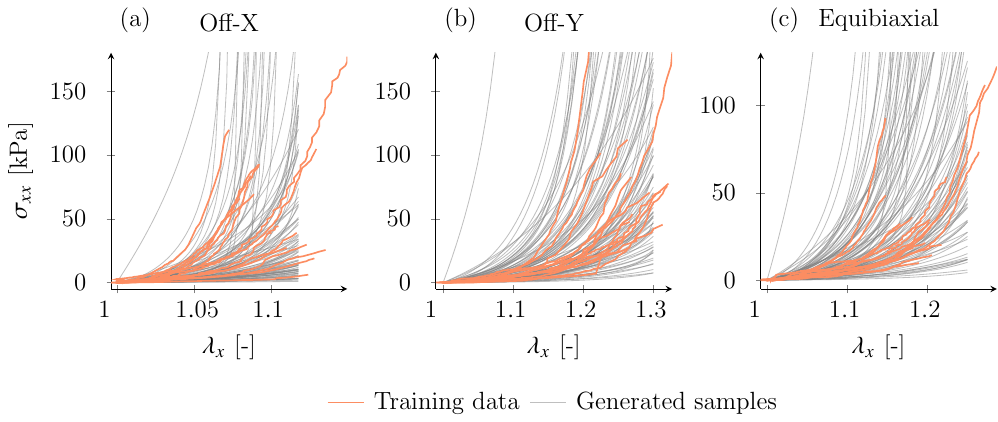}
    \caption{Experimental data of murine skin mechanical response from 15 mice and corresponding samples generated from the diffusion model for Off-X(a), Off-Y (b) and Equibiaxial deformation (c). The generated samples are realistic members of the population of murine mechanical response and satisfy physics constraints \textit{a priori}. }
    \label{fig_mice_stress} 
\end{figure*}

The training data from the 15 mice and the generated samples are shown in Fig. \ref{fig_mice_stress}. The variability in the population is significant which is not surprising for soft tissues such as skin, that show variability between species, with respect to age, sex, anatomical location, and even from one individual to another \cite{luebberding2014mechanical,meador2020}. The data are nonlinear, include anisotropy, and have some noise. Nevertheless, the generated samples from diffusion can realistically describe the behavior of the population. Crucially, because of the structure of the model in Eq. (\ref{eq_strain energy density function_general}), the learned distribution over the strain energy density functions is guaranteed to produce function samples that satisfy the desired physics-constraint.

\begin{figure*}[h!]
    \centering
    \includegraphics[width=\textwidth]{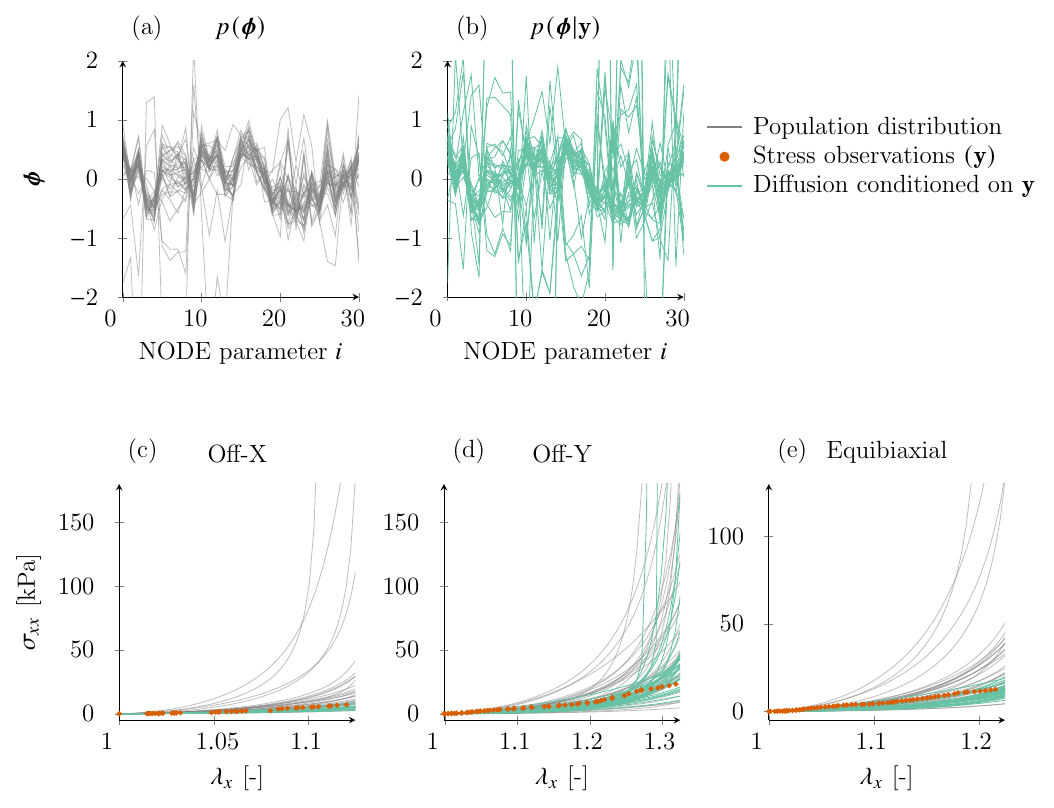}
    \caption{Conditioning strain energy density function generation on experimental stress observations. Once the mechanical behavior of the population has been learned (a), the generation of new strain energy density functions, i.e., new samples $\boldsymbol{\phi}^*$, can be conditioned on stress data $\sigma^i$ by adding a score of the likelihood of this observation to the reverse SDE of the diffusion process (b). The stress-stretch data of the population (grey) as well as of the conditional distribution (orange) based on the stress observations (red) is shown for Off-X(c), Off-Y (d), Equibiaxial (e).  }
    \label{fig_conditional} 
\end{figure*}

With the trained model, one important task is the generation of new strain energy density function samples conditioned on observations. A trivial example is to condition the reverse SDE of the diffusion process on observation of NODE parameters $\bar{\boldsymbol{\phi}}$. An example demonstrating the process of conditioning generation on an observed parameter set is given in the Supplement. This generation process is still useful in that it can be used to produce samples $\boldsymbol{\phi}^*$ similar to existing samples in the training set but with some desired variance $\varsigma$. The more important problem is showcased in Fig. \ref{fig_conditional}, which demonstrates the case of conditioning the generation of NODE parameters $\boldsymbol{\phi}^*$ directly on observed experimental stress data $\boldsymbol{\sigma}^i$. Fig. \ref{fig_conditional} (a) shows generation of parameters from $\hat{p}(\boldsymbol{\phi})$ without any additional observations. The generative model is then conditioned on the (so far unseen) stress data shown in Fig. \ref{fig_conditional} (c-e). The posterior diffused parameters and the stress responses obtained from these parameters are shown with orange lines in Fig. \ref{fig_conditional} (b) and (c-e), respectively. Samples from the population distribution are also shown for comparison. Note that for the conditional distribution we included the score of the likelihood shown in Eq. (\ref{eq_conditional_diffusion_s}) where $\hat{\boldsymbol{\sigma}}(\boldsymbol{\phi})$ is the model that evaluates the stress at a given deformation given the NODE parameters $\boldsymbol{\phi}$. Generation of a material model from the population conditioned on the observation of a quantity of interest of a new individual is of upmost importance in applications such as personalizing a computational model for an individual \cite{lee2021personalized}.

\subsection{Heterogeneous material properties}
\begin{figure*}[h!]
    \centering
    \includegraphics{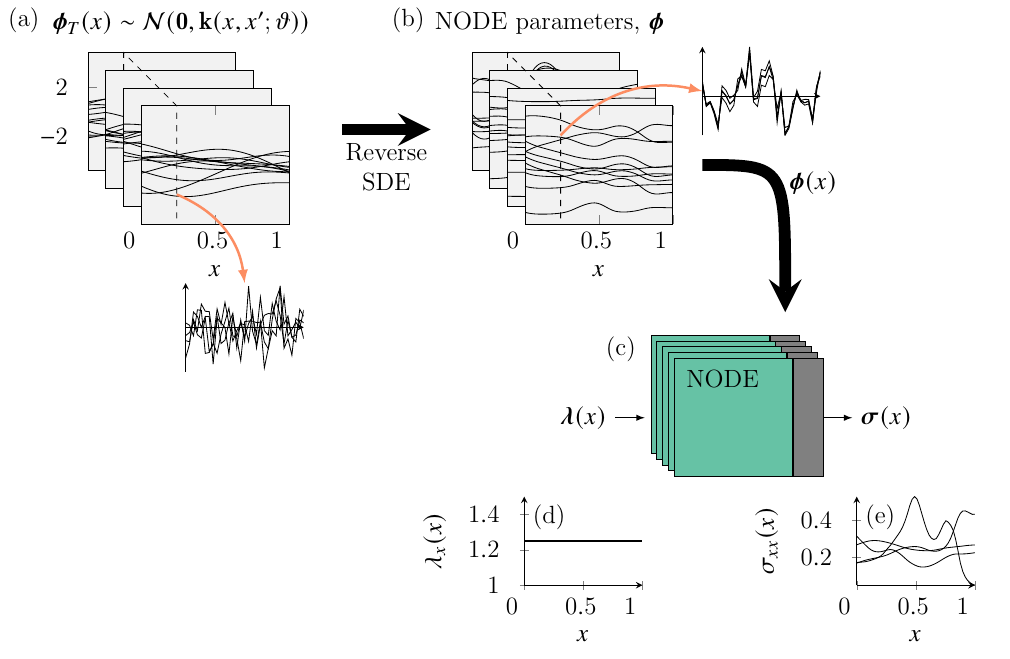}
    \caption{Generation of spatially correlated heterogeneous material fields. Instead of sampling independent standard normal distributions, we sample functions $f(x)$ from a Gaussian process with zero-mean and appropriate covariance function depending on length-scales $\vartheta$ (a). Evaluation of the reverse SDE yields material fields in terms of NODE parameters $\boldsymbol{\phi}(x)$ (b). This enables the evaluation of the NODE constitutive models (c), taking input deformation fields $\lambda_x(x)$ (d) to produce stress fields $\sigma_{xx}(x)$ (e).}
    \label{fig_correlation} 
\end{figure*}

Up to now the reverse SDE has been used to generate point-wise estimates, independent of the spatial coordinates $\mathbf{x}$. However, as described in the Methods section, we extend the reverse SDE so it depends on $\mathbf{x}$ by sampling a GP with zero mean, unit variance, and arbitrary length scales. Generation of spatially correlated parameter fields is shown in Fig. \ref{fig_correlation}. We sample $N_{\text{params}}$ functions of the physical coordinate $x$ from a zero-mean GP to be used as spatially-varying inputs for the reverse SDE (Fig. \ref{fig_correlation} (a)). The reverse SDE then produces spatially-varying NODE parameter sets (Fig. \ref{fig_correlation} (b)). Note that any cross section of Fig. \ref{fig_correlation} (a) resembles a set of samples from the standard normal distribution since the functions $f(\mathbf{x})$ are sampled from a zero-mean GP with unit variance. The diffused parameters are then used in NODE-based material models to predict stresses given stretches over the domain $\mathbf{x}$ (Fig. \ref{fig_correlation} (c)). As an example, we use constant stretch function $\lambda_x(x)=1.25, x\in[0,1]$ and obtain the resulting stress field $\sigma_{xx}(x)$ for a few different samples in Fig. \ref{fig_correlation} (d) and (e). Even though the input for the stress model is a constant field, the output is a spatially-varying stress field $\sigma_{xx}(x)$ with correlation lengths influenced by the correlation of the material properties $\vartheta$.   

\begin{figure*}[h!]
    \centering
    \includegraphics[width=0.9\textwidth]{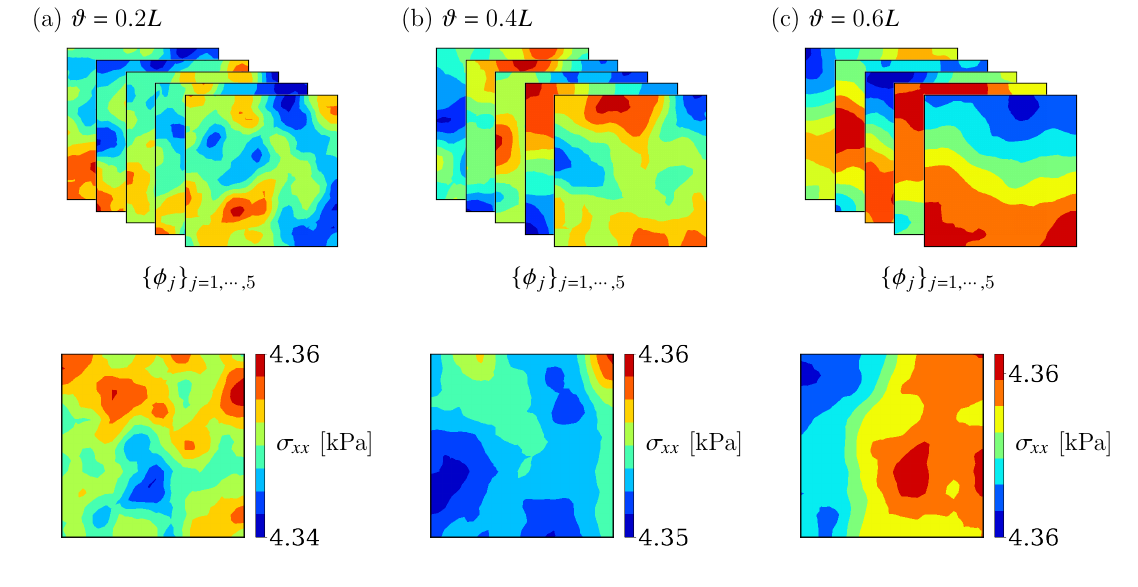}
    \caption{The distribution of some of the sampled parameters $\boldsymbol{\phi}$ generated with the reverse SDE condition on the observation of parameters $\bar{\boldsymbol{\phi}}$ with some variance $\varsigma$. Spatial fields of components $\phi_j$ of the vector $\boldsymbol{\phi}$ plotted on the square domain (top), and the resulting distribution of stiffness (as measured by $\sigma_{xx}$ at $\lambda_x=\lambda_y=1.1$) (bottom) were generated by sampling from three different GPs with length scales; $0.2L$ (a), $0.4L$ (b) and $0.6L$ (c) where $L$ indicates the edge length of the square.}
    \label{fig_square_fields} 
\end{figure*}

\begin{figure*}[h!]
    \centering
    \includegraphics[width=0.9\textwidth]{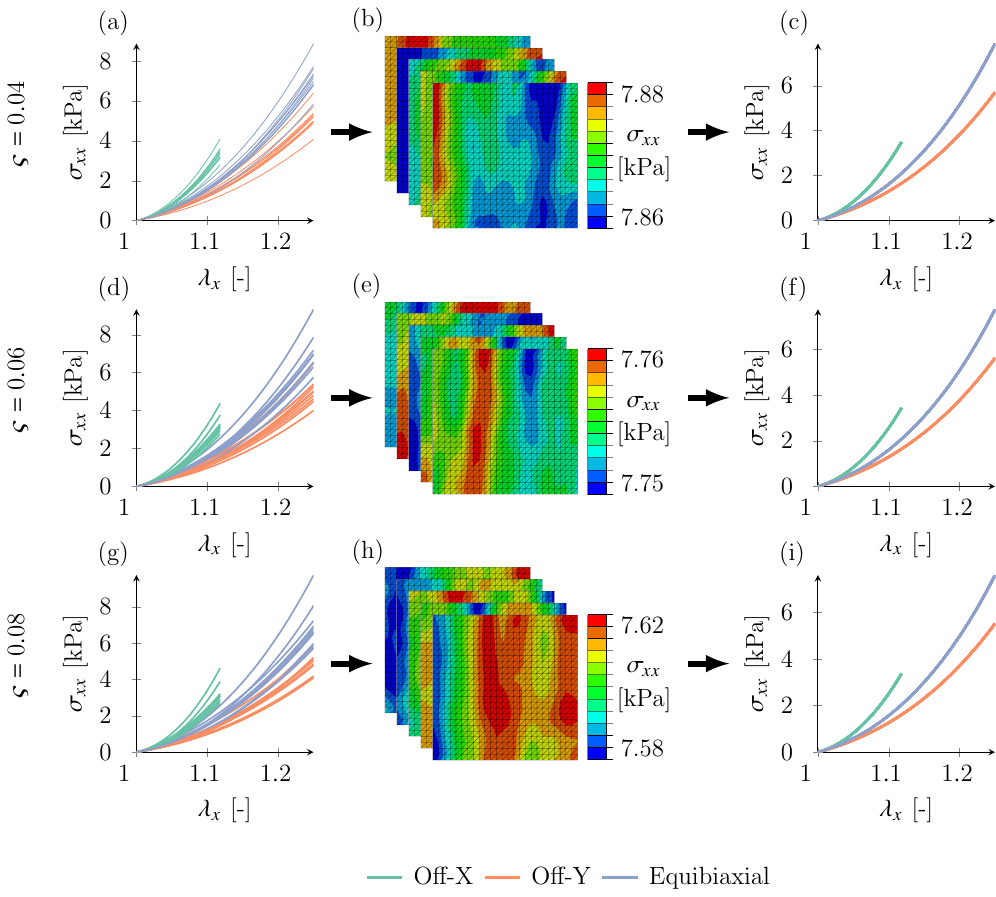}
    \caption{Homogenization of the response of square tissue samples with heterogeneous material properties subjected to off-x, off-y and equi-biaxial deformation. Material behavior close to one of the members of the population were generated by conditioning the diffusion process on observation of the parameters $\bar{\boldsymbol{\phi}}$ of that sample but with different variance $\varsigma$ (a,d,g). The heterogeneous material fields $\boldsymbol{\phi}(\mathbf{x})$ were implemented into the finite element simulations to obtain the heterogeneous strain fields (b,e,h). Integration of the forces on the boundary, together with the stretch boundary condition overlap with the response $(\sigma,\lambda)$ to that defined by the parameters $\bar{\boldsymbol{\phi}}$ regardless of the tissue heterogeneity (c,f,i).  }
    \label{fig_fem_dist} 
\end{figure*}

The first test that we perform is to verify that it is reasonable to use a homogenized response even if the material samples may have spatial heterogeneity. First, we chose one of the individuals from the population and generated samples from the conditional distribution considering we have observed the parameters $\bar{\boldsymbol{\phi}}$ of that individual, as shown in the Supplement Fig.~3. This step generates material responses close to that sample but with some variance $\varsigma$ as illustrated in Figs.~\ref{fig_square_fields} and~\ref{fig_fem_dist}a, d, g. 

Fig.~\ref{fig_square_fields} shows the fields over the square domain of some of the parameters $\boldsymbol{\phi}$ conditioned on the observation of parameters $\bar{\boldsymbol{\phi}}$. We also plot the contour of $\sigma_{xx}$ assuming a \emph{uniform} stretch of $\lambda_x=\lambda_y=1.1$ in the domain. It is important to note that thee fields of $\sigma_{xx}$ \emph{are not} finite element solutions of the stress field but rather a way to visualize the field of material response with a single scalar rather than plotting contours of $\boldsymbol{\phi}$ which have no physical meaning. 

An alternative way of visualizing the material response of the spatial fields in Fig. \ref{fig_square_fields} is to plot the pointwise stress-stretch curves as demonstrated in Fig. \ref{fig_fem_dist}a where it can be seen that the material properties at every point are near the response $\Psi_{\varphi,\bar{\phi}}(\mathbf{F})$ specified by the observed parameters $\bar{\boldsymbol{\phi}}$. Following the generation of heterogeneous skin pieces of tissue, we performed finite element simulations for Off-X, Off-Y and Equi-biaxial deformation in Abaqus using the \texttt{FIELD} command in Abaqus to implement $\boldsymbol{\phi}(\mathbf{x})$ over the mesh, as well as a user material subroutine \texttt{UMAT} to evaluate the NODE-based constitutive model. Heterogeneous strain fields from the finite element simulations are illustrated in Fig. \ref{fig_fem_dist}b, e, h. The variation in the strain field is a consequence of the material field $\boldsymbol{\phi}(\mathbf{x})$ which has characteristic length scale between 0.2 and 0.6 times the edge length. However, despite the stresses being heterogeneous, homogenization of the response (integration of forces over the boundaries) collapses back to the  stress-stretch response  $\Psi_{\varphi,\bar{\phi}}(\mathbf{F})$ defined by parameters $\bar{\boldsymbol{\phi}}$, as illustrated in Fig. \ref{fig_fem_dist}c, f, i. This confirms that even if the square samples of skin tissue were heterogeneous with some length scale $\vartheta$, it is safe to work with the homogenized response of each sample of the population in order to train the diffusion probabilistic model.

\begin{figure*}[h!]
    \centering
    \includegraphics[width=\textwidth]{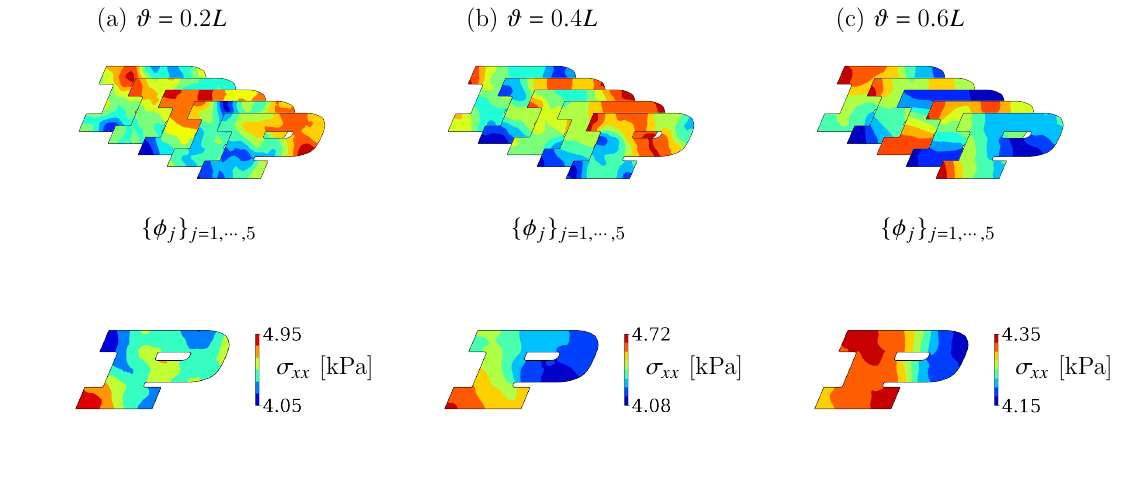}
    \caption{The distribution of (some of the) sampled parameters on the $\mathcal{P}$ (top) and the resulting distribution of stiffness (as measured by $\sigma_{xx}$ at $\lambda_x=\lambda_y=1.1$) (bottom) for three different GP length scales; $0.2L$ (a), $0.4L$ (b) and $0.6L$ (c) where $L$ indicates the width of the $\mathcal{P}$.}
    \label{fig_PP_fields} 
\end{figure*}

Even though the heterogeneity did not affect the overall or homogenized mechanical behavior of square pieces of tissue subjected to biaxial deformation, spatially varying material properties can have an impact on the response of more complex problems which induce stress concentrations. For such problems, heterogeneous material fields can lead to amplification of the maximums stress at regions of interest. We illustrate this phenomenon in the geometry Fig. \ref{fig_PP_fields}.

We again start by generating spatially correlated fields $\boldsymbol{\phi}(\mathbf{x})$ over the geometry $\mathcal{P}\subset\mathbb{R}^2$ shown in Fig. \ref{fig_PP_fields}, where the $\boldsymbol{\phi}(\mathbf{x})$ are  conditioned on the observation of the parameters $\bar{\boldsymbol{\phi}}$ of one of the individuals of the population with some variance $\varsigma$. Contours showing components $\phi_j$ of the vector  $\boldsymbol{\phi}(\mathbf{x})$ over this manifold are illustrated in Fig. \ref{fig_PP_fields}. The components $\phi_j$ were generated with the Mat\'ern kernel Eq. \ref{eq:matern}, with different length scales. To better visualize the mechanical properties with a single scalar contour, we plot the stress field $\sigma_{xx}$, corresponding to a uniform biaxial deformation $\lambda_x=\lambda_y=1.1$. Similar to the heterogeneous materials in the square domain, the contours of $\sigma_{xx}$ in Fig. \ref{fig_PP_fields} are a meaningful way of visualizing the heterogeneous material behavior but are not the solution of linear momentum balance. Finite element simulations with the properties illustrated in Fig. \ref{fig_PP_fields} are explored next.

\begin{figure*}[h!]
    \centering
    \includegraphics{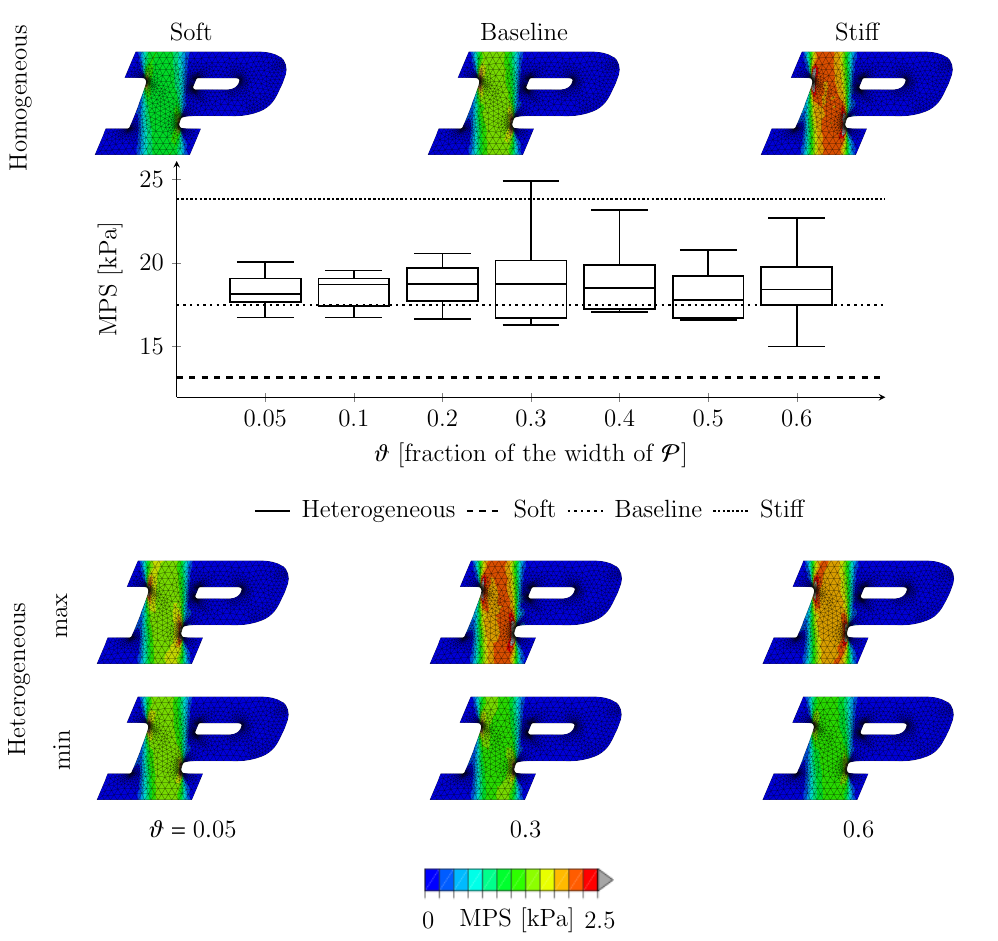}
    \caption{Heterogeneous materials in complex geometries produce maximum principal stresses that are qualitatively different from the uniform material cases. Samples of material properties $\boldsymbol{\phi}$ were generated around the observed parameters $\bar{\boldsymbol{\phi}}$ for either uniform material properties over the domain $\mathcal{P}$ or spatially varying properties $\boldsymbol{\phi}(\mathbf{x})$ over $\mathcal{P}$. The baseline case is defined by material properties $\bar{\boldsymbol{\phi}}$ uniform over $\mathcal{P}$, while soft and stiff  cases are uniform properties but with the lower bound and upper bound of the properties $\boldsymbol{\phi}$ generated around $\bar{\boldsymbol{\phi}}$. Baseline, soft, and stiff MPS contours show stress concentration at corners (top). Heterogeneous fields $\boldsymbol{\phi}(\mathbf{x})$ controlled by the length scale $\vartheta$ produce distribution of MPS with mean always greater than the baseline and confidence interval increasing with $\vartheta$ (middle). Extreme cases of the MPS distribution from the heterogeneous properties are depicted (bottom).  }
    \label{fig_PurdueP} 
\end{figure*}

 Because of the complex geometry, pulling on the top surface  and fixing the bottom surface induces stress concentrations at the corners of the domain, and a band of high stress along the spine of the $\mathcal{P}$ connecting top to bottom surfaces. The uniform material cases are shown in Fig. \ref{fig_PurdueP} top. The most important of the uniform material cases is the \textit{Baseline}, which is when the material behavior corresponds to the parameters $\bar{\boldsymbol{\phi}}$. The two other uniform cases are material responses that bound the samples from the conditional diffusion around $\bar{\boldsymbol{\phi}}$. The \textit{Soft} case is, as the name suggests, the lower bound of the generated $\boldsymbol{\phi}$ around $\bar{\boldsymbol{\phi}}$, while the \textit{Stiff} case is the upper bound from the sampled material responses. The three uniform cases show the same stress distribution, albeit with different magnitude of the maximum principal stress (MPS) from the soft to the stiff cases.

The heterogeneous material distributions $\boldsymbol{\phi}(\mathbf{x})$ over $\mathcal{P}$ are controlled by the length scale $\vartheta$, which is varied from 0.05 to 0.6 times the width of the $\mathcal{P}$. Unlike the heterogeneous square tissue models of Fig.~\ref{fig_fem_dist} for which the response did not depend on the heterogeneity, the heterogeneous $\mathcal{P}$ has a qualitatively different response compared to the uniform case. Observed in Fig. \ref{fig_PurdueP}, middle, the MPS over $\mathcal{P}$ is on average greater than the baseline case for all length scales $\vartheta$. As the length scale increases, there are some samples of  $\boldsymbol{\phi}(\mathbf{x})$ that produce lower MPS than the baseline, approaching the soft uniform case, as well as samples of $\boldsymbol{\phi}(\mathbf{x})$ approaching the stiff uniform case. Indeed, the soft and stiff uniform cases can be seen as samples $\boldsymbol{\phi}(\mathbf{x})$ when the length scale goes to infinity. What is more interesting is the distribution of MPS for smaller length scales. For $\vartheta<0.2$ the MPS distribution is always greater than the baseline case while still bounded by the stiff uniform case. This implies that heterogeneous material properties cannot be ignored for complex geometries. Extremes of the MPS distribution for different $\vartheta$ are depicted in  Fig.~\ref{fig_PurdueP}, bottom.

\section{Discussion}
We presented a framework for generating constitutive models for hyperelasticity that \textit{a priori} satisfy desirable physics constraints. The model is based on diffusion and specifically score matching models. One key idea that we introduced is to guarantee polyconvexity of random samples by working with NODE-based strain energy functions that are polyconvex by design \cite{tac2022node}. Thus, rather than learning distribution over  a discrete set of strain energy density function evaluations directly, the framework is akin to hypernetworks \cite{krueger2017bayesian}: we estimate the probability density of NODE parameters with diffusion given stress-strain data. We showcase the success of the framework in learning the distribution of material response both when presented with synthetic data and experimental data from murine skin. Once trained, the generative model can be used to sample new material models from the population, or models conditioned on new experimental observations. Another key idea is the extension of the generative framework to sample spatially correlated fields in order to model heterogeneous materials on complex geometries. Material heterogeneity can lead to amplification of stress concentrations. Thus, among other examples, applications involving heterogeneous materials such as skin and biological tissues requires sampling spatially correlated fields in order to guide decision making under uncertainty. 

Usual application of diffusion-based generative models are inherently finite-dimensional, e.g., images \cite{yang2022diffusion}. There have been some recent efforts focused on generative models for functions or manifolds. For example, Du et al. \cite{du2021learning} and Dupont et al. \cite{dupont2022data} represent functions with MLPs and propose to learn a latent space and a hypernetwork to go from the latent space to the weights and biases of the MLPs. Diffusion probabilistic models are then used on latent space samples. Instead of parameterizing functions, it is possible to work directly with finite-dimensional data by sampling both the domain and the function values, as shown in \cite{zhuang2022diffusion,dutordoir2023neural,elhag2023manifold}. However, sampling domain inputs and corresponding function values makes it more difficult to impose constraints on the functions. One approach is to add the desired constraints as a likelihood in the reverse SDE \cite{chung2022improving}. Our approach extends previous work in a manner more similar to hypernetworks, but with a network architecture that \textit{a priori} satisfies the function constraints we are interested in. One design choice is which weights and biases should be allowed to be random variables and which should be fixed parameters, or whether to introduce a latent space as in \cite{dupont2022data}. We found that varying the last layer of the NODE was flexible enough to capture the variability in the population. The rest of the weights and biases were shared among all members of the population, therefore providing the flexibility needed to capture the average response. The common weights and biases were treated as fixed parameters and estimated directly by minimizing a standard training loss. The approach was effective for hyperelasticity, but it is possible that more complex behavior, e.g. viscoplasticity, might require more sothisticated methods to avoid the curse of dimensionality. 


Tissues have inherent variability in mechanical properties from one individual to another and even spatially within the same individual \cite{staber2018random,hauseux2018quantifying}, skin being a good example of this phenomenon \cite{jor2013computational,Joodaki_review}. Clinically, uncertainty in the mechanical and biological response of skin has an impact on the resulting stress distributions after reconstructive surgery \cite{lee2018}, or tissue expansion \cite{lee2020}, to name a couple examples. Thus, any clinical decision making based on computational models of skin needs to account for the variability across individuals as well as heterogeneous material properties. The same is true for other tissues \cite{kouznetsova2001approach}, and even beyond, for structural materials with similar heterogeneity \cite{li2019predicting,matouvs2017review}. One of the challenges of trying to do uncertainty analysis for skin or other tissues is that the uncertainty might be too large to improve decision making with respect to current \textit{standard of care}. Conditional generation of material responses given some observation for a new individual is an effective strategy to improve computational models for clinical settings. Noninvasive techniques to estimate skin properties \textit{in vivo} have emerged over the past decade \cite{mueller2021reliability,laiacona2019non,liang2009biomechanical}. However, data from these tests is usually not enough to fully characterize the material response \cite{song2022bayesian,mueller2021reliability}. Our framework enables estimation of the distribution of mechanical response across a population based on detailed mechanical test data, with the capacity to update the generation of new samples given new, simpler observations. We anticipate that this framework can help in clinical decision making by generating models of skin for a new individual based on abundant prior knowledge of the population that is captured by our generative model.

For a given sample of a material, modeling spatial heterogeneity requires generation of spatially correlated fields. We showed that the reverse SDE of the diffusion process can be extended to generate these fields over complex domains. The key idea in standard diffusion models is that generation of new samples can be done starting from the standard normal distribution. We extended the reverse SDE by sampling functions $f(\mathbf{x})$ from a Gaussian process with zero mean and unit variance in the kernel Eq. (\ref{eq:kernel}). Sampling from a GP for rectangular domains is trivial and can be done with many existing packages such as GPy \cite{erickson2018comparison}. For complex geometries, we follow the eigen-expansion based on the Laplacian operator outlined in \cite{costabal2022delta,elhag2023manifold}. One key question is whether or not the Euler-Maruyama scheme of the spatially correlated fields converges or not. A numerical investigation showing convergence of the stochastic PDE \ref{eq_Euler_stochastic_PDE} with respect to to number of samples, time step, and spatial resolution is provided in the Supplement.

Using the heterogeneous material fields we found that spatial heterogeneity had no impact on overall response of square pieces of tissue under biaxial loading. Thus, we can conclude that for many \textit{ex vivo} testing scenarios with uniform boundary conditions, it is safe to work with the homogenized stress-strain data to train the generative model \cite{you2022physics}. For complex tissues, e.g. three-dimensional solids such as myocardium or ligaments \cite{kakaletsis2021,estrada2020mr}, the full-field strain is needed in order to calibrate the model. For example, we show that for a toy problem with a \emph{P} geometry, material heterogeneity amplifies stress concentrations with respect to the uniform material case, and this amplification depends on the length scale of the heterogeneity. Future work should focus on the inverse problem of learning the length scale and material properties given the full-field strain and boundary conditions.    

This work is not without limitations. The framework is focused on hyperelasticity but tissues can have more complex mechanical behavior such as viscoelasticity, damage, fracture, etc. \cite{zhang2022effects,holzapfel2017modeling}. We have previously extended the data-driven framework based on NODEs to capture viscoelasticity \cite{tacDatadrivenAnisotropicFinite2023}. Ours is also not the only data-driven method for constitutive modeling of tissues \cite{fuhg2022learning,vlassis2020geometric,linka2023new}. It remains to test whether the generative framework can be easily extended to the more complex behaviors, and whether other architectures besides NODE yield similar results to what we show here. Another limitation is that we work with homogenized response and not full-field strain data. We show that this is adequate for the type of \textit{ex vivo} tests we are concerned with, but would not apply to three-dimensional tissues with non-uniform boundary conditions. Thus, additional work is needed to set up the probabilistic modeling framework that directly deals with heterogeneous strain fields during density estimation. The third limitation to point out is the limited experimental data. For murine skin we had stress-strain response from 15 mice. However,  as seen with the synthetic example, which is based on a popular material model \cite{may1998constitutive}, up to 100 samples might be needed to estimate the material model across a population. Nevertheless, the synthetic example also shows that even with limited data the framework can estimate reasonable parameter distributions. These results underscore the need for more tissue stress-strain data repositories with very clear protocols and annotations in order to enable the proper training of probabilistic models. 

\section{Conclusions}

We have presented a novel framework for generative hyperelasticity. Our framework satisfies physical constraints by construction, such as polyconvexity and objectivity, which enables us to run finite element simulations with convergence guarantees. Combined with diffusion probabilistic models, our method is able to capture the typical variability observed in the experimental response of materials such as soft tissues. By itself, this is a useful result that can be used for uncertainty quantification in simulations. We take the approach one step further by creating spatially correlated, but heterogeneous, material properties in a fully data-driven way. We observe that the length scale of the correlation has a major effect on the probability distribution of the maximum stress obtained, which could lead to different design or clinical decisions. Overall, we believe this a major step forward to make data-driven methods useful in engineering practice.

\section{Acknowledgements}
VT and IB acknowledge the support of AFOSR under the grant number FA09950-22-1-0061.
FSC and MR acknowledge the support of the Open Seed Fund of the School of Engineering at Pontificia Universidad Cat\'olica de Chile. ABT acknowledges support from National Institute of Arthritis and Musculoskeletal and Skin Diseases, National Institute of Health, United States under award R01AR074525.

\section{Supplementary material}
All data, model parameters and code associated with this study are available in a public Github repository at \url{https://github.com/tajtac/node_diffusion}.


\end{document}